# Observation of a phonon bottleneck effect on the thermal depopulation from a photo-excited shallow defect in silicon

Sergio Revuelta[1], Hai I. Wang[2,3], Mischa Bonn[2] and Enrique Cánovas[1,*]

[1]Instituto IMDEA Nanociencia, Campus Universitario de Cantoblanco, Faraday 9, 28049, Madrid, Spain

[2]Max Planck Institute for Polymer Research, Ackermannweg 10, 55128, Mainz, Germany

[3]Nanophotonics, Debye Institute for Nanomaterials Science, Utrecht University, Utrecht, The Netherlands

[*]enrique.canovas@imdea.org

ABSTRACT:

We report the observation of a phonon bottleneck effect impacting the thermal depopulation of photo-excited shallow defects in high-resistivity silicon. Using time-resolved terahertz (THz) spectroscopy, near-bandgap excitation produces a pronounced temporal delay in photoconductivity, indicating that a fraction of photogenerated charge carriers are temporarily trapped immediately after excitation. By analyzing the frequency-resolved complex photoconductivity as a function of pump-probe delay and photon energy, we attribute this delay to the presence of a localized shallow state situated approximately 40 meV from the band edge, which competes with silicon's indirect band-to-band absorption. The zero-order kinetic profile of the temporal delay, its invariance with respect to photon flux, and its temperature dependence collectively support the existence of a phonon bottleneck that hinders the thermal release of electrons from this shallow trap. To our knowledge, this represents the first experimental evidence of a phonon bottleneck effect associated with the thermal activation of shallow traps in photoexcited silicon. These findings provide new microscopic insight into carrier relaxation dynamics in silicon and highlight the significance of electron-phonon interactions in the ultrafast processes governing materials used in optoelectronic applications.

## I.    INTRODUCTION

The interplay between electrons and phonons is a pivotal aspect in the study of many-body physics, with important implications in both, fundamental and applied research [1,2]. When photons are absorbed in a semiconductor, electrons are promoted from the valence band towards the conduction band. If the absorbed photons have an energy which is larger than the semiconductor's bandgap onset, the photo-excited "hot" carriers will relax to the band edge, dissipating their excess energy into the lattice by phonon emission [3,4]. When phonons with the proper energy and momentum are eventually not available (e.g., when depleted at high laser fluences), the hot carrier relaxation is consequently slowed down. This so-called phonon-bottleneck effect has been widely studied in bulk semiconductors [4–6] and more recently in low-dimensional nanostructures [7–9], where the mismatch between electronic energy levels and the available phonon population was expected to hinder the efficient cooling of hot carriers. In both reported areas, the bottleneck is linked to hot carrier cooling (i.e., the release of excess energy from a hot electron). Here, we present what is, to our knowledge, the first demonstration of a pho7non-bottleneck effect associated with the thermal depopulation of a photoexcited shallow defect state in a semiconductor—specifically, an energy gain process wherein a trapped electron is released via phonon absorption.

We experimentally observe that near-bandgap photoexcitation of a high-resistivity float-zone silicon wafer results in a temporal delay in the rise of the time-resolved photoconductivity, as monitored by time-resolved terahertz (THz) spectroscopy [10,11]. Analyzing the frequency-resolved complex photoconductivity vs. pump-probe delay, we conclude unambiguously that the temporal delay is related to a temporal change in carrier density rather than mobility. This demonstrates that part of the photoexcited charge carriers are temporarily localized right after being photo-generated, and then "released" with a temporal delay. By tuning the energy of the pump above and below silicon's onset of absorption we observe that the strength of the resolved temporal delay is photon energy-dependent, reaching a maximum when exciting the samples with ~1.08 eV photons, slightly below the bandgap energy of 1.12 eV. From this analysis, we conclude that a localized state placed ~40 meV below (or above) silicon's conduction (valence) band competes favorably with the absorption of photons by the indirect band-to-band process. The linear lineshape obtained for the dynamics linked with the temporal delay right after photo generation suggests that a zero-order process governs the thermal release of trapped charge carriers. We attribute these observations to a phonon bottleneck effect for the thermal release of electrons (holes) from the shallow trap towards silicon's conduction (or valence) band. The analysis of the temperature-dependent dynamics, and the invariance with respect to photon flux for a given photon energy, support the notion that a phonon bottleneck effect on the thermal depopulation of a photoexcited shallow defect in silicon is responsible for the observed singular dynamics.

## II.    EXPERIMENTAL

The sample analyzed in this work consisted of a 0.5-mm thick high-resistivity silicon float zone wafer with <100> orientation (Sigma-Aldrich ID: 646687, $\rho$=100-3000 Ω cm). A Ti:Sapphire amplified laser system providing 800 nm of wavelength output at 1 KHz and ~50 fs of pulse width was employed to run the optical pump-terahertz probe experiment [11,12]. The THz radiation was generated via optical rectification on a ZnTe crystal of 0.5 mm thickness. The detection was done via electro-optical sampling on a ZnTe crystal of identical characteristics. For optical excitation, we used a commercial optical parametric amplifier (SpectraPhysics TOPAS-prime) powered by the femtosecond laser system (see Supplemental Material [13] and reference [14] for further details).

## III.    RESULTS

In Figure 1, we present the time-resolved optical-pump terahertz-probe (OPTP) dynamics for the real part of the photoconductivity, Re[Δσ(t)], when the sample is excited near its bandgap onset with 1.12 eV photons [15] (at 14.1 μJ/cm²). The real part of the photoconductivity Re[Δσ(t)] = $N(t) \cdot e \cdot \mu$; where $N(t)$ is the density of photogenerated charge carriers, $e$ is the electron charge and $\mu$ is the carrier mobility).  Immediately after photoexcitation, at time zero on the *x*-axis, there is a sharp rise in the signal, linked to the almost instantaneous photogeneration of charge carriers in the sample. Notably, a noticeable "slow" dynamic component emerges right after this initial sub-ps sharp rise. The slow rise accounts for ~40 percent of the total OPTP amplitude and reveals a photo-induced process that lasts up to ~70 ps (Figure 1a). Note that exciting the sample well above its bandgap (see Figure S1 in SI) does not reveal the slow dynamic component, as expected for a silicon sample where conductivity occurs right after the photo generation of long-lived (typically hundreds of μs)  free charge carriers.

To understand the nature of the singular temporal response when exciting the sample near its bandgap, we disentangle the mobility and carrier density contributions to the time-resolved photoconductivity as a function of the pump-probe delay. In Figure 1b, we show measurements of the frequency-resolved complex photoconductivity spectra at 2, 9, 29, and 54 ps after photoexcitation. Filled and open symbols represent the experimentally retrieved real and imaginary components, respectively. Here, all the results, independently of pump-probe delay, can be very well described using the Drude model:

$$\Delta\tilde{\sigma} = \frac{\omega_p^2 \varepsilon_0 \tau}{1 - i\omega\tau} \qquad \text{(Eq. 1)}$$

Where $\omega_p$  is the plasma frequency, $\tau$ is the scattering time, $\omega$ is the angular frequency and $\varepsilon_0$ is the vacuum dielectric constant. The best fits to the Drude model are shown in Figure 1b, represented by solid and dashed lines for the real and imaginary components of the frequency-resolved photoconductivity, respectively. The fits provide an invariant scattering rate upon pump-probe delay, with a median value of 156±8 fs, and a rise in the plasma frequency from 6.5(±0.1)·10¹¹ Hz at 2 ps to 9.5(±0.1)·10[11] Hz at 54 ps. Figure 1c summarizes the results arising from the fitting protocol in a wide range of different pump-probe delays (see Figure S2 in the Supplemental Material [13] for the complete set of frequency-resolved complex photoconductivities). Here the filled symbols represent the inferred squared plasma frequencies ($\omega_p^2$) and open symbols the charge carrier's scattering times ($\tau$). As evident from the plots, the scattering time is largely invariant to the pump-probe delay, demonstrating that the energy-dependent slow rise resolved in OPTP data is not related to a change in charge carrier mobility. In line with this conclusion, the temporal evolution of the square of the plasma frequency as a function of pump-probe delay nicely resembles the dynamics resolved in the OPTP trace (see Figure 1c top panel), underlining that the time-dependent ingrowth in conductivity results purely from an increasing carrier density. Strictly speaking, a change in the plasma frequency could be related either to a change in the carrier density, $\Delta N$, or to a change in the effective mass of the carriers, $m^*$. However, as we are exciting silicon at the bandgap energy, directly to the bottom of the conduction band, a variation in the effective mass seems unlikely. In this respect, an increasing carrier density with pump-probe delay indicates that a fraction of the absorbed photons generate species that are transiently localized (i.e., they have null mobility).

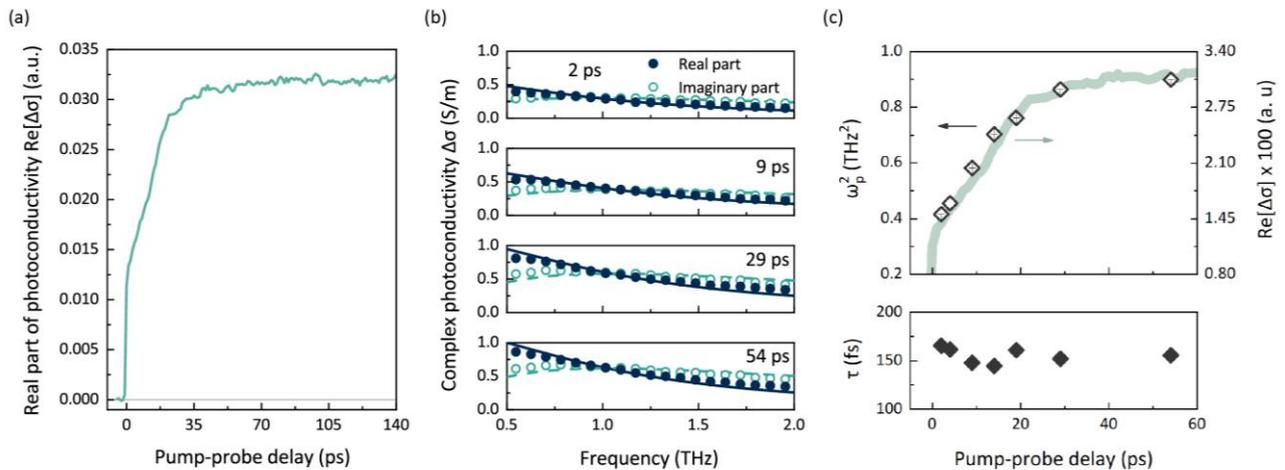

**Figure 1.** (a) Time dependence of the real photoconductivity at 1.12 eV of photon-pump energy and 14.1 μJ/cm² of photon flux. (b) Exemplary measurements of the frequency-resolved complex photoconductivity retrieved at indicated pump-probe delays. Filled and open symbols refer to the real and imaginary components of the complex photoconductivity, respectively. Solid blue and dashed green lines indicate the best Drude fit to the experimental data. (c) Top panel: square of the plasma frequency, $\omega_p^2$, compared with the real part of the photoconductivity shown in Figure 1a; Bottom panel: scattering time, τ, retrieved from the best Drude fits to the frequency-resolved complex photoconductivity as a function of the pump-probe delay.

We interpret our results as evidence that a portion of the photogenerated charge carriers become temporarily localized in a defect state, which is subsequently depopulated through phonon-mediated processes. To test this hypothesis, we investigated how the observed temporal delay in carrier generation varies with the incident photon energy, spanning from 1.55 eV (800 nm) to 0.95 eV (1310 nm). Figure 2a shows the normalized real part of the photoconductivity as a function of the photon-pump energy (see Figure S3 and Table S1 in the Supplemental Material [13]). The results reveal that as the photon pump energy is decreased from 1.37 eV (900 nm) to 1.08 eV (1150 nm), the "slow" rise component becomes increasingly prominent, reaching its maximum at 1.08 eV. Notably, when the pump energy is further reduced to 0.95 eV (1310 nm), this delayed feature in the OPTP dynamics disappears almost entirely. We note here that Drude fit analysis to all the frequency-resolved complex conductivity data corroborates that the OPTP slow rise seen in all the data sheets is related to a change in charge carrier density and not with a change in charge carrier mobility independently of the photon energy employed (for details see Figure S4-S6 and Table S2 in the Supplemental Material [13]). By analyzing the amplitude of the OPTP trace at 2 ps after photoexcitation, we resolve a Gaussian-like feature centered at 1.08 eV (see Inset in Figure 2a and Table S3 in the Supplemental Material [13]). We interpret this feature as indicating a shallow state located ~40 meV below (above) the conduction (valence) band for the analyzed sample. A discussion of the eventual chemical identity of this shallow state will be given later.

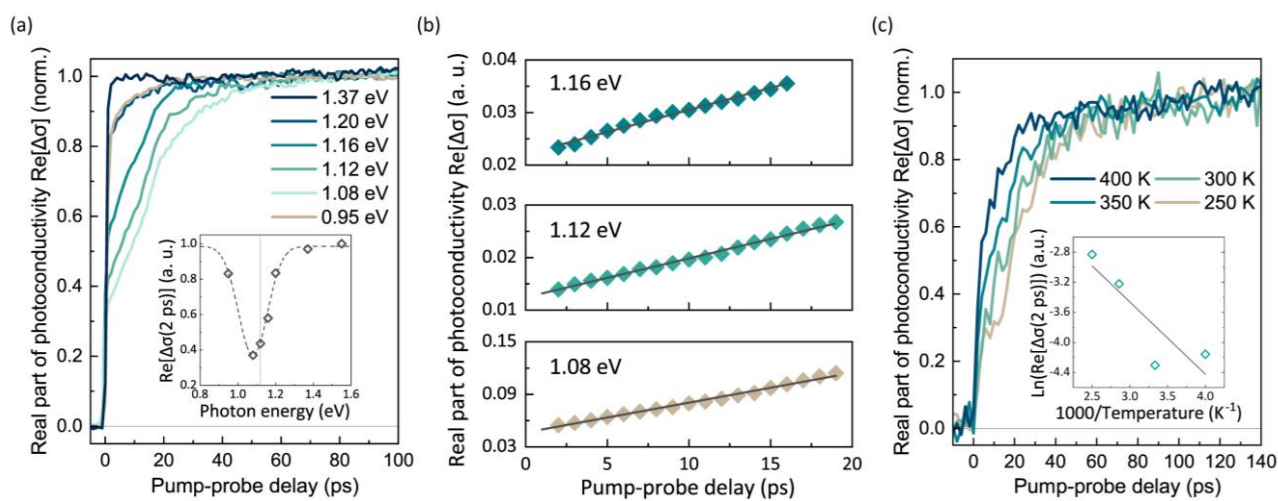

**Figure 2.** (a) Normalized time dependence of the real photoconductivity at different photon-pump energies as indicated. Inset: Normalized real photoconductivity at 2 ps after photo-injection as a function of the pump-photon energy. Vertical line refers to the silicon band gap energy of 1.12 eV. The dashed line is a Gaussian fit to the experimental points. (b) Time-dependent amplitude of the real part of the photoconductivity following excitation with 1.16, 1.12, and 1.08 eV photon energy. Solid lines are a linear fit to the experimental data points. (c) Normalized real part of the photoconductivity vs pump-probe delay measured at 1.12 eV of photon-pump energy at a fluence of 25 μJ/cm$^2$ at the indicated temperatures (see Figure S9 in the Supplemental Material [13] for the non-normalized traces). Inset: Logarithm of the amplitude signal 2ps after photo-injection as a function of the temperature's inverse. Linear line refers to a linear fit to the experimental data.

To gain further insight into the observed temporal delay associated with the shallow level at 1.08 eV, we examine the evolution of the slow-rise component in greater detail. If the delay were governed by a standard Arrhenius-type thermal emission from a shallow trap, it would typically be reflected in the dynamics as a first-order, exponential response [16,17]. Instead, our measurements indicate that a zero-order process better fits the data (see Figure 2b), as evidenced by the initial linear-like behavior of the transient photoconductivity observed when the sample is excited with various photon energies (refer to Discussion S2 and Figure S7 in the Supplemental Material [13]). Zero-order kinetics are characteristic of scenarios where a saturated, rate-limiting step dominates the process [18]. This observation leads us to conclude that our results are a manifestation of a phonon bottleneck effect in the thermal release of carriers from the experimentally resolved shallow state. If this scenario is correct, the process should be limited by the number of available phonons and not depend on the initial concentration of free carriers (given by the density of impinging photons). In fact, by measuring the OPTP traces at different fluences, we record an invariant modulation of the real part of the photoconductivity (see Figure S8 in the Supplemental Material [13]), consistent with our expectations. To further investigate this aspect, we performed temperature-dependent measurements to resolve the quenching of the effect when increasing the sample temperature [19]. Figure 2c shows that this is the case, as the amplitude for the delayed photogeneration for a given photon energy is quenched when increasing the sample temperature from 250 to 400 K (see Figure S9 for the non-normalized traces in the Supplemental Material [13]). The activation energy of the system can be process can be resolved from the slope of the logarithmic representation of the real part of the photoconductivity just after photoexcitation (i.e., within the first 2 ps) vs the inverse of the temperature. This relationship is described by the Arrhenius equation [20] (see Inset of Figure 2c, Table S4 in the Supplemental Material [13] and reference [21] for details). From the above-mentioned analysis we obtain an apparent activation energy of ~82 meV. This value exceeds the maximum phonon energies available in silicon, which show an onset ~65 meV [22], an aspect that supports the notion that the observed delay in the photogeneration may be associated with a phonon bottleneck effect.

The results, discussion, and analysis provided above, demonstrate that photo-generated electrons occupy transiently a shallow level in the forbidden gap [23], and that this level is thermally depopulated suffering a transient bottleneck effect induced by a lack of phonons with the proper energy/momentum. While at present we are unable to conclude the specific nature of the shallow defect, in the following we speculate about its possible origin. The specifications given by the wafer provider indicate an unintentional n-doping character (no dopants added) for the high resistivity FZ sample. In this respect, we should in principle rule out that phosphorus (P) doping (which provides a shallow donor with ~40meV energy) being behind the resolved bottleneck. To further check this scenario, we analyzed another wafer from the same provider but doped with P (Sigma-Aldrich ID: 647780, $\rho$= 1 - 10 $\Omega$ cm, $N_D$ ~ 1·10$^{15}$ cm$^{-3}$). Notably, the n-type P-doped wafer does not show the zero-order kinetic fingerprint under the same experimental conditions (see Figure S10 in the Supplemental Material [13]). As such, we need to assume the existence of grown-in defects in the high-resistivity wafer being present in the as-received samples [24]. A second look at the specifications given by the wafer provider indicates a rather large oxygen and carbon content for the analyzed FZ wafers (Oxygen content: ≤ 1~1.8 x $10^{18}$ /cm³; Carbon content: ≤ 5 x $10^{16}$ /cm³), which might indicate on-purpose addition during growth [25]. In this respect, and following the literature, we tentatively suggest that the observed peak at 1.08 eV might be related to shallow donors located ~40 meV below the CB that have indeed been related to the presence of C-O complexes and also with C-O complexes containing nitrogen [26,27]. More work is needed to ascertain the nature of these defects, which is currently underway in our group. In any case, we would like to highlight here that, independently of the origin of the resolved state at 1.08 eV, we believe our main observation holds, which is the demonstration of a phonon bottleneck for the thermal depopulation of this state upon near-bandgap photoexcitation in the analyzed silicon wafer ( a diagram illustrating the process is shown in Figure S11 in the Supplemental Material [13]).

## IV.    CONCLUSIONS

In summary, in this paper, we have investigated the ultrafast photogeneration of charge carriers in a semi-insulating silicon wafer using time-resolved THz spectroscopy. The observation of a delay in the photo generation of free charge carriers upon photon absorption near silicon´s bandgap onset is linked to the transient thermal depopulation of a shallow level located at 40 meV below (above) the CB (VB). The temporal delay is rationalized assuming a phonon-bottleneck effect for the thermal depopulation of this shallow level. This scenario is supported by the line shape of the recorded dynamics, their temperature and pump-fluence dependence, as well as by pump-energy dependent measurements. In contrast to the abundant literature on bottlenecks induced during hot carrier cooling, to our knowledge, our results reveal, for the first time, a phonon bottleneck effect on the thermal depopulation from a photo-excited, below-bandgap, shallow defect in silicon.

**Acknowledgements:**

This publication has been financially supported by the grants TED2021-129624B-C44 and PID2023-148369OB-C44 funded by the Spanish Research Agency MCIN/AEI/10.13039/501100011033 and by the European Union "NextGenerationEU"/PRTR. IMDEA Nanociencia acknowledges support from the 'Severo Ochoa' Programme for Centres of Excellence in R&D (MINECO, Grant SEV-2016-0686). We also acknowledge support from the Erasmus+ program and from the Max Planck Society. We appreciate L. Di Virgilio for helpful discussions and support with measurements.

# Supplemental Material

## Supplemental Discussion S1

**Optical pump-terahertz probe.** We record the time-dependent real part of the photoconductivity by monitoring the evolution of the THz transmitted waveform as a function of time after photoexcitation, at the maximum peak of the THz pulse in the time-domain. Being recorded in the time-domain, it represents a summatory of sinusoidal components for all frequencies retrieved in the frequency-domain. This protocol is done to maximize the signal to noise ratio. Furthermore, the maximum in the time-domain is where the derivative is zero and it is on the limit of small differential signals related to absorption and phase shifts, so it represents primarily changes in the absorbance rather than phase shifts [1].

The frequency-resolved complex photoconductivity can be inferred by measuring the transmitted THz pulses from the non-excited sample, $\vec{E}_r$, and from the pump-induced change in the absorption, $\Delta\vec{E}$. If the optical excitation is nonzero over the whole sample (i.e., weak optical absorption where $\alpha L < 1$, where $\alpha$ is the optical absorption and L is the sample's thickness), it is possible to obtain a quasi-uniform excitation in the sample and obtain an expression that reads [2]:

$$\Delta\tilde{\sigma} = -\frac{n+1}{Z_0 L}\frac{\Delta\vec{E}(\omega)}{\vec{E}_r(\omega)} \qquad (\text{eq. S1})$$

Where $L$ is the sample thickness, $Z_0$ is the intrinsic impedance of free space and $n$ is the refractive index of the unexcited medium.

## Supplemental Discussion S2

**Process order for a phonon-bottleneck.** The process of photogeneration in an indirect semiconductor as silicon when photoexciting with a photon energy below the direct transition, implies we need a photon and a phonon to get a free charge-carrier. That is, we can understand the system a process involved by two reactants (photons and phonons) and one product (free charge-carriers). As the real part of the photoconductivity its defined as $\Delta\sigma = N \cdot e \cdot \mu$ (where N is the charge-carrier density, e is the electron charge and $\mu$ is the charge carrier mobility), and our results of the frequency-resolved complex photoconductivity (see Figure 2 in the main text and Figures S4, S5 and S6 in the Supplemental Material) resolved a change in the plasma frequency (i.e., charge-carrier density) and not in the scattering time (i.e., charge-carrier mobility), the variations in $\Delta\sigma$ can only be attributed to changes in the free charge-carrier density.

In a first approximation, the order of a process can be studied by representing the process in different scales and determine whether it is obtained a straight line. A zero-order process is typically obtained in reactions when there is a limitation of one of the reactants and its characterized by a linear response in the representation of [A] vs t, where [A] is the concentration of the product. A first-order process can be obtained when plotting Ln[A] vs t. A second-order process on the other hand its typically obtained if 1/[A] vs t gives a linear tendency [3].

As it can be seen in Figure S7, we always obtain a linear tendency during the first ps for the cases of irradiating the silicon sample under 1.16, 1.12 and 1.08 eV of photon energy. This could imply we are facing a phonon-bottleneck effect happing during the photogeneration of free charge-carriers. This is also corroborated by the fact that by changing the photon flux we did not obtain a different lineshape in the photoconductivity (see Figure S8).

**Supplemental Figure S1**

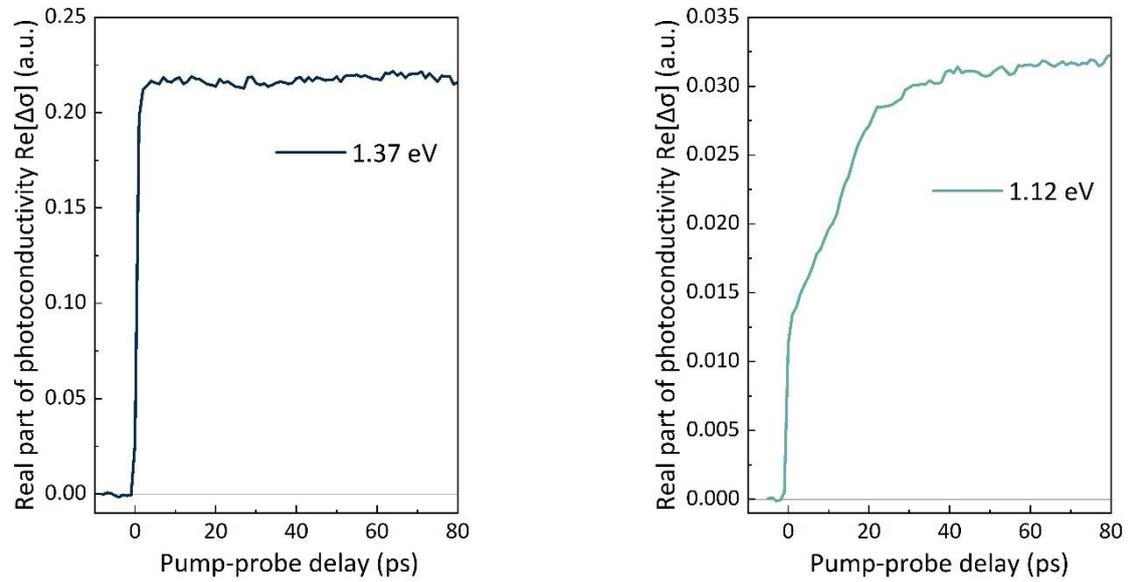

**Figure S1**. Real photoconductivity as function of the pump-probe delay at different photon energies as indicated. Fluences used: 4.5 µJ/cm$^2$ at 1.37 eV (900 nm) and 14.1 at 1.12 eV (1100 nm).

**Supplemental Figure S2**

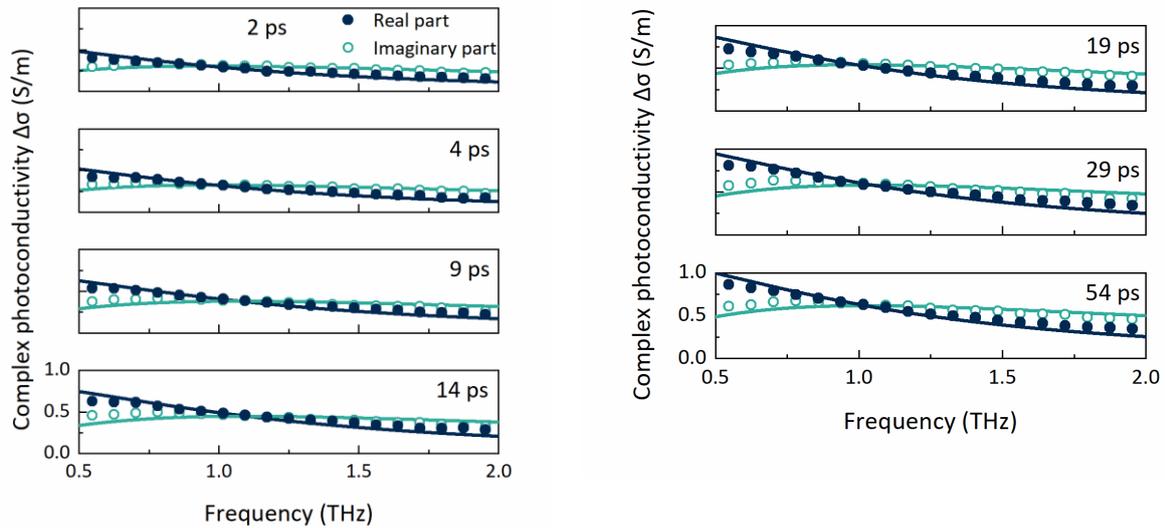

**Figure S2.** Frequency-resolved complex photoconductivity retrieved at different pump-probe delays at 1.12 eV of photon energy. Filled and open symbols refer to the real and imaginary components of the complex photoconductivity respectively. Solid blue (real) and green (imaginary) lines indicates the best Drude fit to the experimental data.

**Supplemental Figure S3**

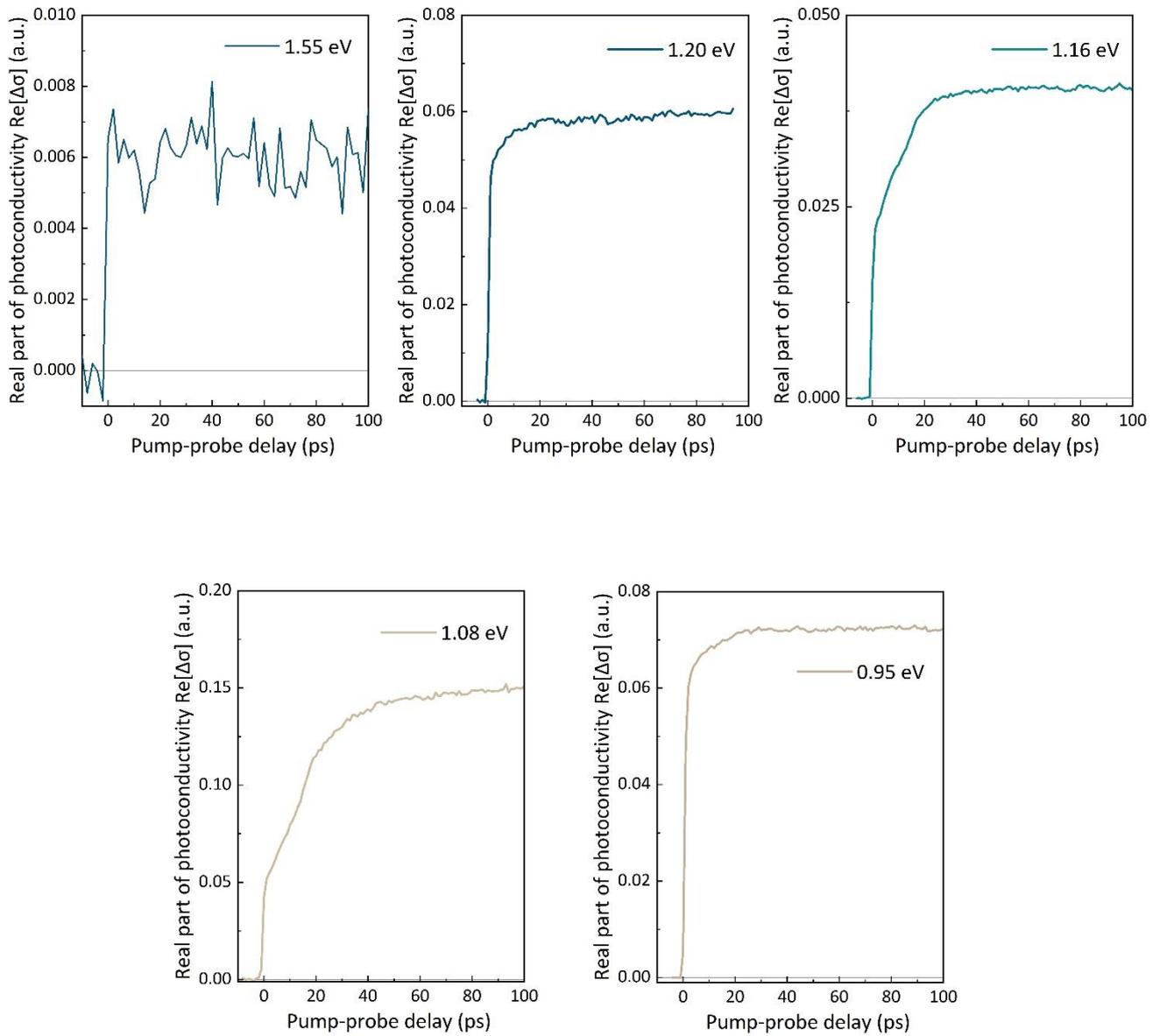

**Figure S3**. Real photoconductivity as function of the pump-probe delay at different photon energies as indicated.

**Supplemental Figure S4**

To resolve the frequency-resolved complex photoconductivity obtained under this photon energy (1.55 eV) we cannot use the expression given in the main text of the article (eq. 1), as the photoexcitation is not homogeneous along all the bulk material. In this case, as the material has a strong optical absorption (~12 μm of optical absorption [4]), we are generating the charge-carriers in a very thin slab near the surface. Notice that the thickness of the material is 0.5 mm). In such case, we can use the Tinkham approximation to resolve the complex photoconductivity as [1]:

$$\Delta\tilde{\sigma} = -\frac{n+1}{Z_0 d}\frac{\Delta\tilde{E}(\omega)}{\tilde{E}_r(\omega)} \text{ (eq. S2)}$$

Where $d$ is the penetration depth of the optical excitation, $Z_0$ is the intrinsic impedance of free space and $n$ is the refractive index of the unexcited medium.

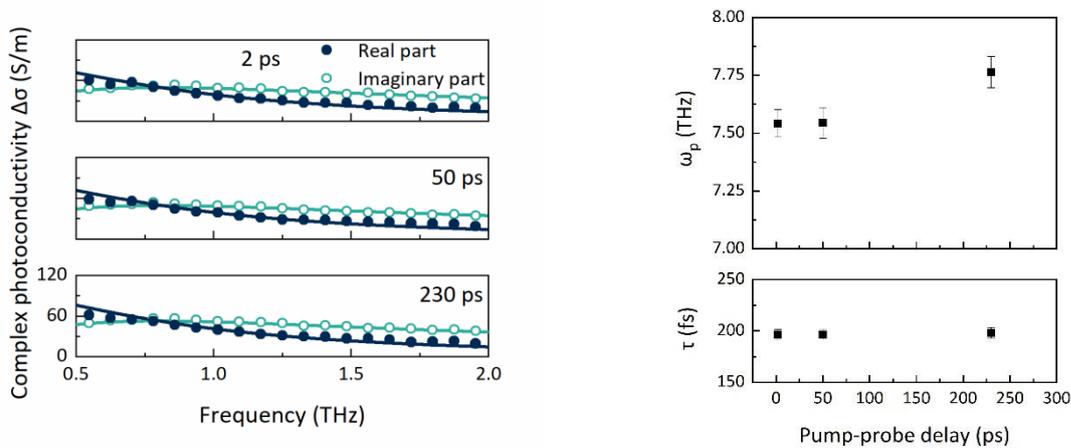

**Figure S4.** On the left: Frequency-resolved complex photoconductivity retrieved at different pump-probe delays at 1.55 eV of photon energy. Filled and open symbols refer to the real and imaginary components of the complex photoconductivity respectively. Solid blue (real) and green (imaginary) lines indicates the best Drude fit to the experimental data. On the right: Plasma frequency, $\omega_p$, (top panel) and scattering time, $\tau$, (bottom panel) retrieved from the best Drude fits to the frequency-resolved complex photoconductivity as a function of pump-probe delay.

**Supplemental Figure S5**

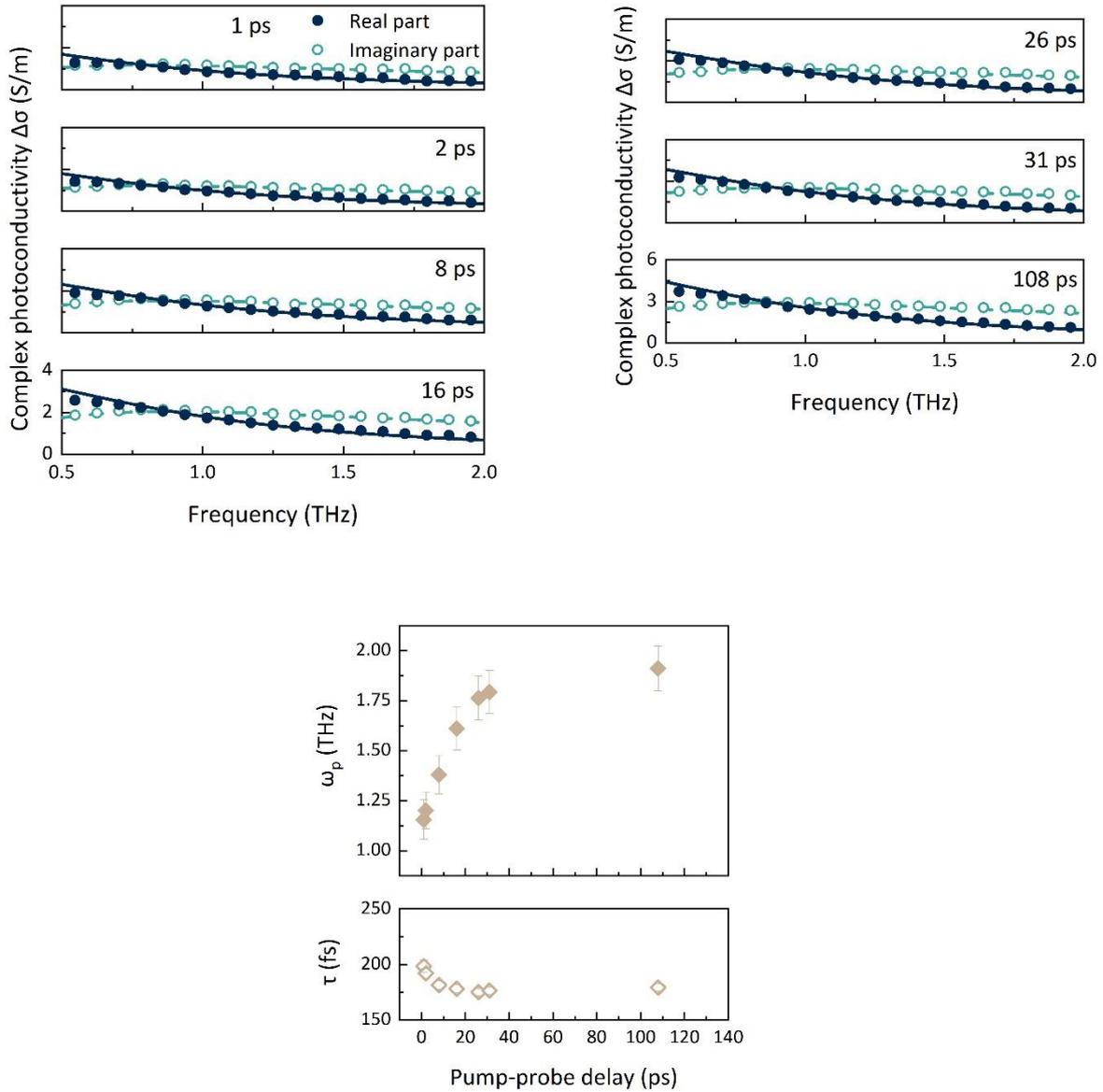

**Figure S5.** On the top: Frequency-resolved complex photoconductivity retrieved at different pump-probe delays at 1.08 eV of photon energy. Filled and open symbols refer to the real and imaginary components of the complex photoconductivity respectively. Solid blue (real) and green (imaginary) lines indicates the best Drude fit to the experimental data. On the bottom: Plasma frequency, $\omega_p$, and scattering time, $\tau$, retrieved from the best Drude fits to the frequency-resolved complex photoconductivity as a function of pump-probe delay.

**Supplemental Figure S6**

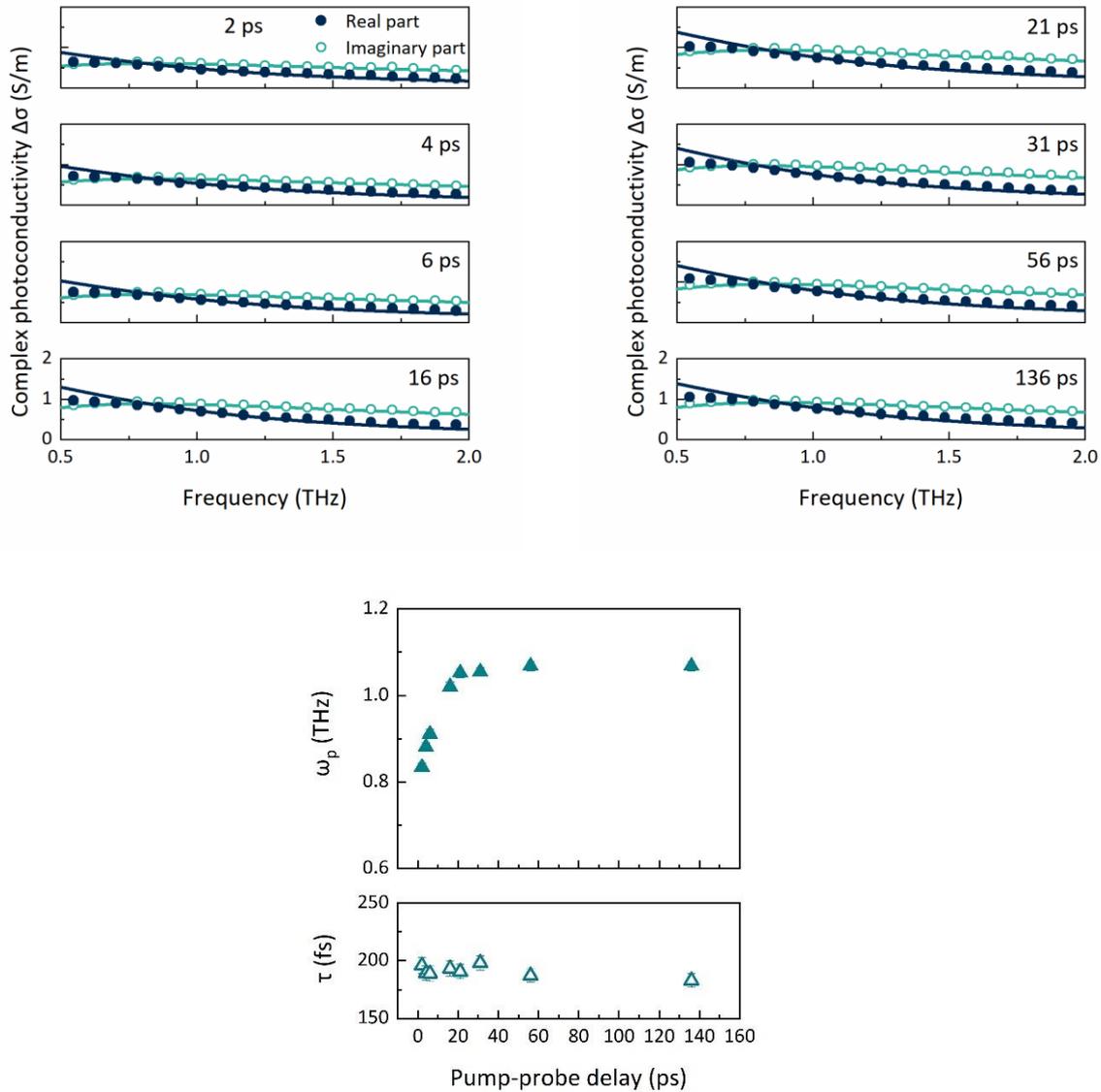

**Figure S6.** On the top: Frequency-resolved complex photoconductivity retrieved at different pump-probe delays at 1.16 eV of photon energy. Filled and open symbols refer to the real and imaginary components of the complex photoconductivity respectively. Solid blue (real) and green (imaginary) lines indicates the best Drude fit to the experimental data. On the bottom: Plasma frequency, $\omega_p$, and scattering time, $\tau$, retrieved from the best Drude fits to the frequency-resolved complex photoconductivity as a function of pump-probe delay.

**Supplemental Figure S7**

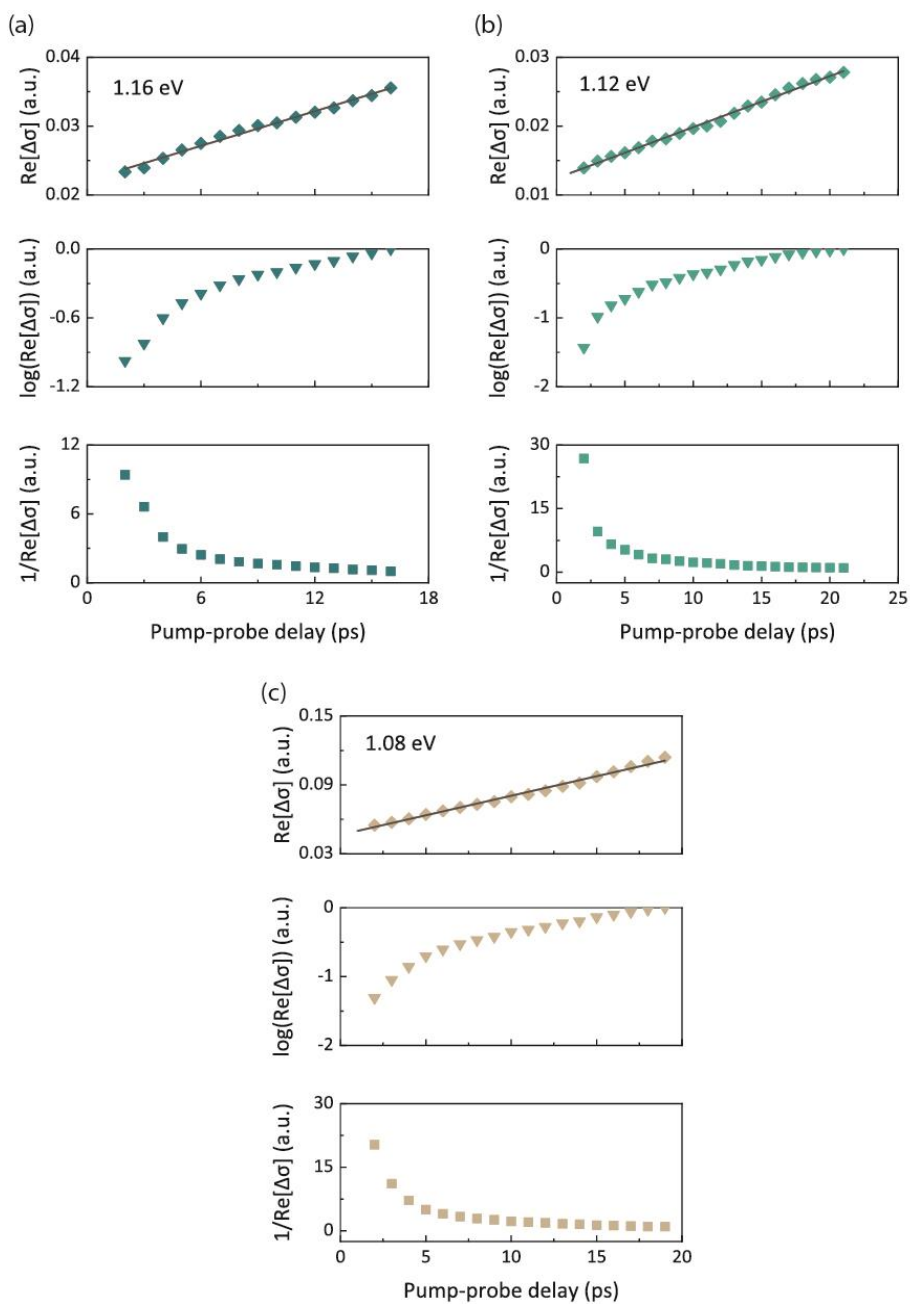

**Figure S7.** Real part of the photoconductivity represented in different scales for the photon energies of 1.16 (a), 1.12 (b) and 1.08 eV (c). Top panels: Real part of the photoconductivity vs time, $[\Delta\sigma]$ vs t. Middle panels: Ln$[\Delta\sigma]$ vs t. Bottom panels: $1/[\Delta\sigma]$ vs t. A straight line when representing $[A]$ vs t is a possible indication of a zero-order process, typically obtained when there is a limitation of one of the reactants. The obtained slopes of the linear fits are $(8.4\pm0.2)\cdot10^{-4}$, $(7.4\pm0.1)\cdot10^{-4}$ and $(3.4\pm0.1)\cdot10^{-3}$ for the data at 1.16, 1.12 and 1.08 eV respectively.

**Supplemental Figure S8**

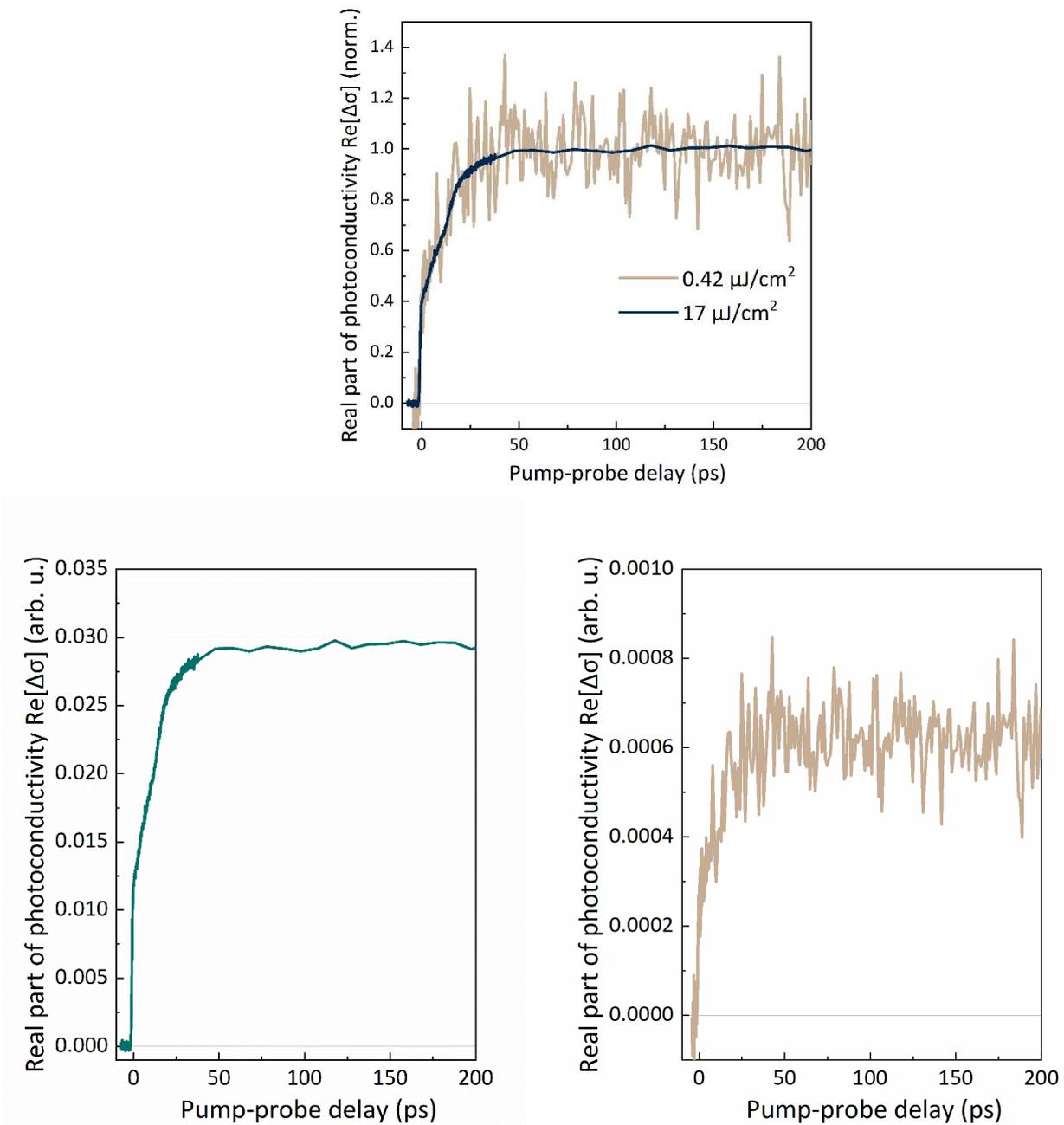

**Figure S8.** On top: Normalized real part of the photoconductivity at 1.12 eV of photon-pump energy at two different fluences of 17 and 0.42 μJ/cm². On bottom: Real part of the photoconductivity vs pump-probe delay measured at 1.12 eV of photon energy (1100 nm) at a fluence of 17 (left panel) and 0.42 μJ/cm² (right panel).

**Supplemental Figure S9**

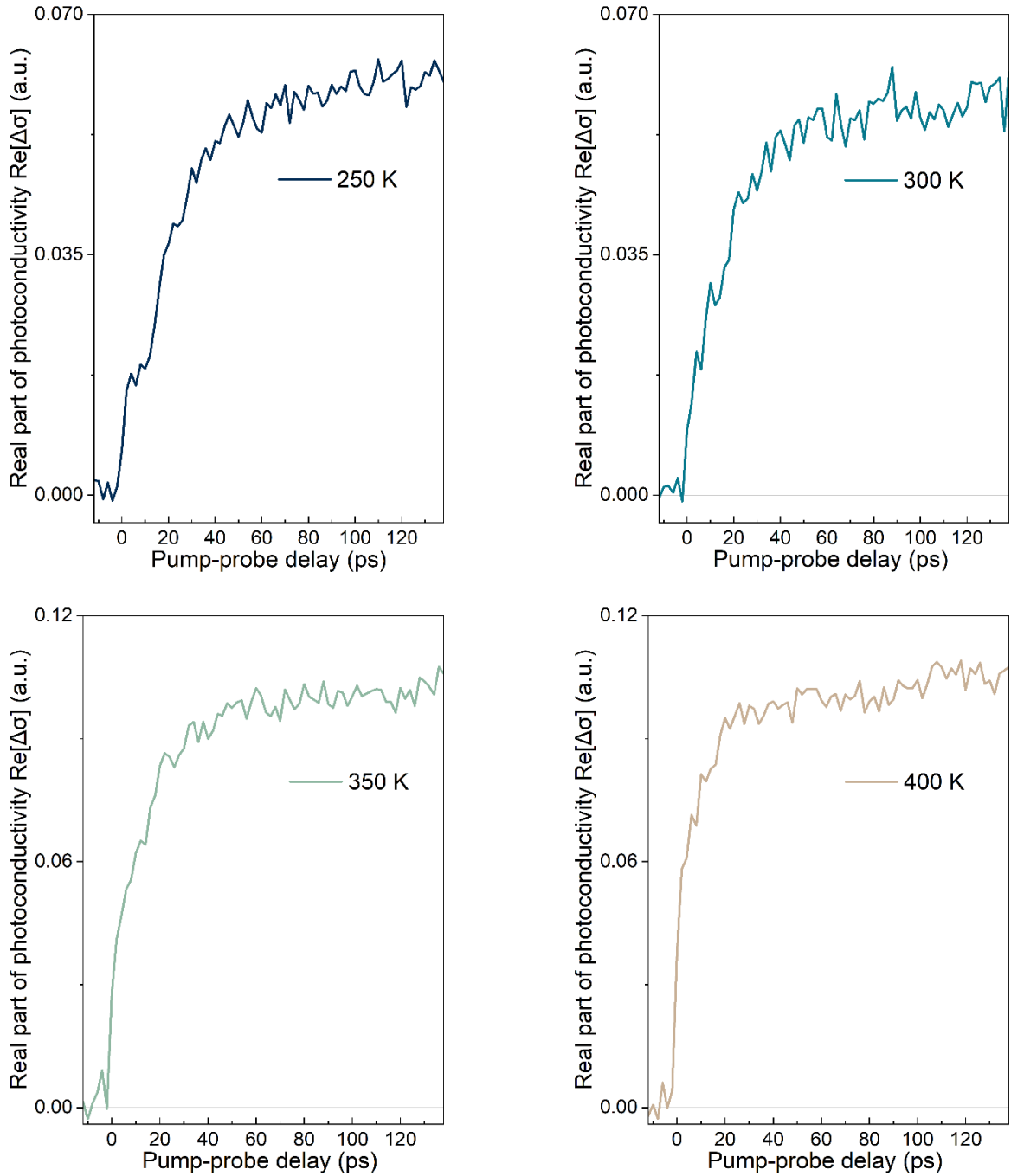

**Figure S9**. Real photoconductivity as function of the pump-probe delay at different temperatures as indicated at 1.12 eV of photon energy at 25 μJ/cm².

**Supplemental Figure S10**

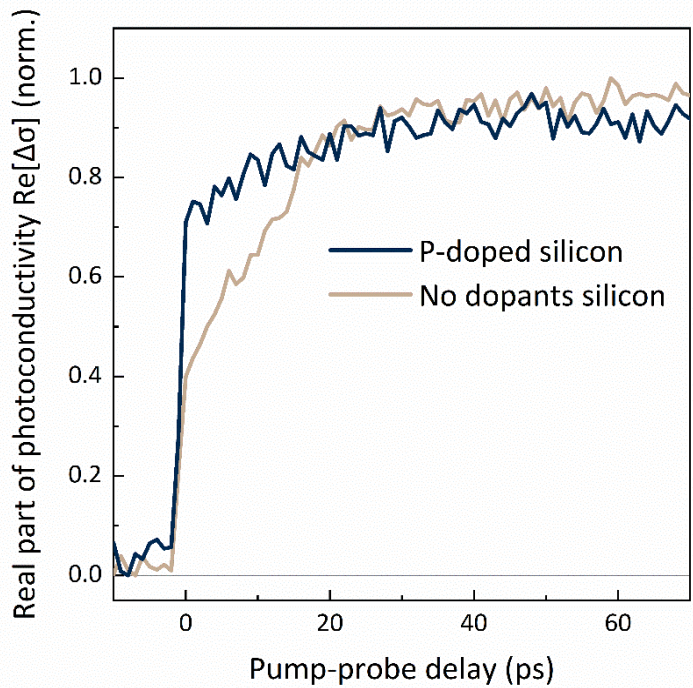

**Figure S10.** Normalized real part of the photoconductivity vs pump-probe delay measured at 1.12 eV of photon energy (1100 nm) at a fluence of 12.1 µJnJ/cm$^2$ for two distinct silicon samples with and without N dopants. Resistivity rage of doped wafer between 7.5 and 12.5 Ω cmΩcm (N$_D$ ~ 1·10$^{15}$ cm$^{-3}$).

**Supplemental Figure S11**

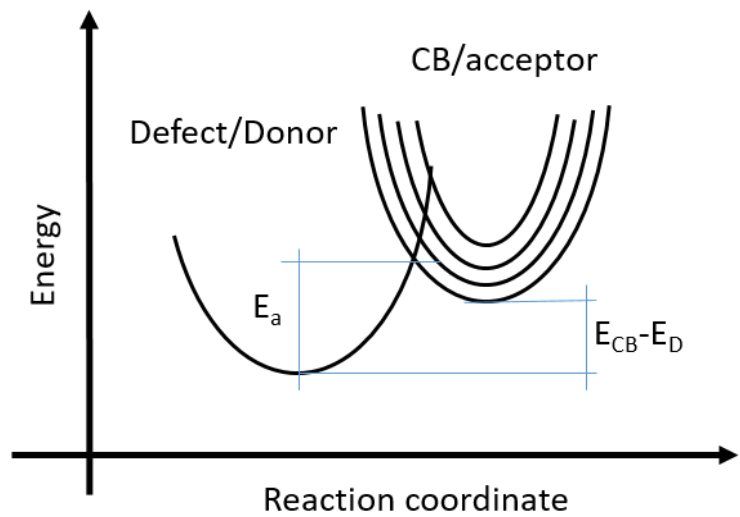

**Figure S11.** Reaction coordinate diagram illustrating the phonon-bottleneck mechanism. Near-bandgap excitation promotes electrons from the valence band to a defect donor ($E_D$) localized below de CB; the thermal release of electrons trapped in $E_D$ towards the CB requires phonons with an energy equal to $E_a$, where $E_a > E_{CB}-E_D$. A bottleneck occurs due to a limitation in the number of available phonons required for this process to take place. This phonon-bottleneck provoke a delay in the photoconductivity at early times.

**Supplemental Table S1**

In order to increase the signal-to-noise ratio when needed, we have changed the fluence at some photon-pump energies when possible. Table S1 shows the fluence used for each photon-pump energy.

**Table S1.** Summary of fluences used in Figure 1 and Figure 3a at the different photon-pump energies.

| Wavelength (nm) | Photon energy (eV) | Fluence |
|---|---|---|
| 800 | 1.55 | 0.28 μJ/cm$^2$ |
| 900 | 1.37 | 4.5 μJ/cm$^2$ |
| 1030 | 1.20 | 21.3 μJ/cm$^2$ |
| 1070 | 1.16 | 20.5 μJ/cm$^2$ |
| 1100 | 1.12 | 14.1 μJ/cm$^2$ |
| 1150 | 1.08 | 44.4 μJ/cm$^2$ |
| 1310 | 0.95 | 9.1 μJ/cm$^2$ |

**Supplemental Table S2**

The silicon penetration depth at the used photon energies have been calculated following the reported data of M. A. Green for an intrinsic silicon wafer at RT [4].

**Table S2.** Summary of the equations and penetration depths used.

| Pump-wavelength (nm) | Photon-pump energy (eV) | Abs. coeff. (cm$^{-1}$) | Penetration depth (μm) | αL | Eq. used |
|---|---|---|---|---|---|
| 800 | 1.55 | 850 | 12 | 42.5 | S1 |
| 900 | 1.37 | 303 | 33 | 15.15 | - |
| 1030 | 1.20 | 30.2 | 331 | 1.51 | - |
| 1070 | 1.16 | 8 | 1250 | 0.4 | 1 |
| 1100 | 1.12 | 2.7 | 3700 | 0.135 | 1 |
| 1150 | 1.08 | 0.68 | 14700 | 0.034 | 1 |
| 1310 | 0.92 | $2.7\cdot10^{-5}$ | $3.7\cdot10^{8}$ | $1.35\cdot10^{-6}$ | - |

**Supplemental Table S3**

Below can be found the equation and the values obtained from the Gaussian fit to the experimental data points obtained at 2 ps after photoexcitation at the different photon pump energies.

$$y = y_0 + A \cdot e^{-\frac{(x-x_c)^2}{2\omega^2}} \quad \text{(eq. S3)}$$

**Table S3.** Parameters of the Gaussian fit represented in Inset Figure 2a.

| $y_0$ | $x_c$ | $\omega$ | $A$ |
|---|---|---|---|
| $0.98 \pm 0.01$ | $1.08 \pm 0.01$ | $0.08 \pm 0.01$ | $-0.62 \pm 0.02$ |

**Supplemental Table S4**

We have resolved the activation energy from the slope of the logarithmic representation of the Re[Δσ] vs the inverse of temperature at the pump-probe delay of 2 ps (i.e., just after photoinjection) using the Arrhenius equation [5]. The raw data can be seen in Figure S9.

From the Arrhenius equation (eq. S4):

$$k = A \cdot e^{-E_a/RT} \quad \text{(eq. S4)}$$

Where $k$ is the reaction rate, $A$ is the Arrhenius factor, $E_a$ is the activation energy, $R$ is the universal gas constant (8.314 J mol$^{-1}$ K$^{-1}$) and $T$ is the temperature. From the logarithmic plot of the Re[Δσ] vs the temperature, the slope is related to $E_a/R$. The activation energy then can be converted from KJ/mol to eV with the Avogadro number (N$_A$ = 6.022 ... $\cdot 10^{23}$ $mol^{-1}$):

$$1 \; eV = 1.602 \cdot 10^{-22} \; KJ * N_A = 96.485 \; KJ \; mol^{-1} \quad \text{(eq. S5)}$$

**Table S4.** Parameters of the linear fit ($y=mx+n$) represented in Inset of Figure 2c.

| $m$ | $n$ | Activation Energy (KJ/mol) | Activation Energy (meV) |
|---|---|---|---|
| --960±385 | -0.58±1.2 | 7.98±3.21 | 82.70±33.22 |

We have selected a pump-probe delay of 2ps for this calculation to avoid any problems linked to the IRF of our setup, where the THz probe has a temporal width of approx.. 1.8ps. In any case, we have calculated as well the activation energy for 0ps pump-probe dealy, obtaining an activation energy of ~105meV. A figure that is also consistent with the proposed bottleneck.

**Supplemental References**